\documentclass{IEEE_lsens}

\usepackage{textcomp}

\usepackage{cite}

\usepackage[T1]{fontenc} % optional enhanced font encoding
\usepackage{amsmath}
\usepackage[cmintegrals]{newtxmath}

\usepackage{bm}

\usepackage{array}

\usepackage{url}
\ifCLASSINFOpdf
\else
\fi
\providecommand{\hypersetup}[1]{\relax}

\hypersetup{pdftitle={Bare Demo of IEEE\_lsens.cls for IEEE Sensors Letters},%<!CHANGE!
pdfsubject={Typesetting},%<!CHANGE!
pdfauthor={Wolfgang Schmailzl},%<!CHANGE!
pdfkeywords={Class, IEEE, IEEE\_lsens, IEEE Sensors Letters, LaTeX, Typesetting, TeX}}%<^!CHANGE!}

\usepackage{siunitx}

\usepackage{BOONDOX-cal}
\def\BibTeX{{\rm B\kern-.05em{\sc i\kern-.025em b}\kern-.08em
    T\kern-.1667em\lower.7ex\hbox{E}\kern-.125emX}}

\DeclareUnicodeCharacter{202F}{\,}

\RequirePackage{graphicx}
\graphicspath{ {./images/} }
\usepackage{subcaption}

\usepackage{placeins}

\usepackage{dblfloatfix}

\begin{document}

\markboth{Vol.xx, No.~xx, October~2020}{PREPRINT}

% article subject line 
%\IEEELSENSarticlesubject{Sensor Applications}

\title{Tip Avalanche Photodiode - A new generation Silicon Photomultiplier based on non-planar technology}

\author{Eugen Engelmann, Wolfgang Schmailzl, Peter Iskra, Florian Wiest, Elena Popova\\ and Sergey Vinogradov, \IEEEmembership{Senior Member, IEEE}
\thanks{\textit{Corresponding authors: Eugen Engelmann and Wolfgang Schmailzl}}
\thanks{E. Engelmann, F. Wiest and P. Iskra are with KETEK GmbH, Munich, Germany (e-mail: eug.engelmann@gmx.de, flw@ketek.net, pis@ketek.net).}
\thanks{W. Schmailzl is with the Institute of Physics, Universität der Bundeswehr München, Munich, Germany (e-mail: wolfgang.schmailzl@gmail.com).}
\thanks{E. Popova is with the National Research Nuclear University MEPhI, Moscow, Russia (e-mail: elenap73@mail.ru).}
\thanks{S. Vinogradov is with the National Research Nuclear University MEPhI, Moscow, Russia and with the P. N. Lebedev Physical Institute, Moscow, Russia (e-mail: vinogradovsl@lebedev.ru).}
}

% Manuscript received line
\IEEELSENSmanuscriptreceived{Submitted to IEEE sensors October 14, 2020;}

\IEEEtitleabstractindextext{%
\begin{abstract}
The Silicon Photomultiplier (SiPM) is a mature photodetector concept that is applied in a variety of applications ranging from medical imaging to automotive LiDAR systems. Over the last few years, improvements of the sensor performance are gradually approaching to a saturation. In this work we present our new concept to overcome the intrinsic limitations of planar configurations of electrodes.
Our non-planar technology is based on focusing and enhancing the electric fields by tip-like electrodes. The shape of the electric field and the lack of typical micro-cell edges, allows us to exclude cell separation boundaries and eliminate dead space around active cell areas. Our design provides a high-density micro-cell layout with a high geometric efficiency. It resolves the well-known trade-off between the detection efficiency and the dynamic range.
The first "Tip Avalanche Photodiode" (TAPD) prototypes show a remarkable geometric efficiency above 80\,\% for a micro-cell pitch of 15\,µm. This directly translates into a photon detection efficiency (PDE) record peak value of 73\,\% at 600\,nm with respect to the state-of-the-art SiPMs. Moreover, the PDE remains above a value of 45\,\% up to a wavelength of 800\,nm with another record value of 22\,\% at 905\,nm. The reduced micro-cell capacity allows for a fast recovery time below 4\,ns, which improves the operation at high photon rates.
Overall, the TAPD is anticipated to be a very promising SiPM generation for various wide-spectral and high-dynamic-range applications in health science, biophysics, particle physics and LiDARs.
\end{abstract}

}

\IEEEpubid{This work has been submitted to the IEEE for possible publication. \\
Copyright may be transferred without notice, after which this version may no longer be accessible.}

\maketitle

\section{Introduction}
A majority of modern Silicon Photomultipliers (SiPMs) is designed as an array of independent Single-Photon Avalanche Diode (SPAD) cells with individual quenching resistors connected to a common electrode. An active photosensitive part of the cell is formed by planar pn-junctions. The inactive cell periphery is typically formed by guard rings or trenches to provide an independent operation of the cells in the Geiger breakdown processes. This design has been initially developed in Russia \cite{A,B,1}. Since the mid-2000s, the planar SiPM design concept has been utilized with some modifications by Hamamatsu \cite{2}, SensL \cite{C}, STM \cite{3}, FBK \cite{4}, Excelitas \cite{5}, and KETEK \cite{6}. It was worldwide recognized as the SiPM, a new photon-number-resolving avalanche detector of outstanding performance \cite{7}.

Nowadays, the planar SiPMs are approaching the physical limits of the design performance with its main inherent trade-off between photon detection efficiency (PDE) and dynamic range (DR) because of the dead space at the cell periphery \cite{Vinogradov2011}. 
This trade-off becomes challenging when developing a SiPM which should be sensitive for a wide spectral range, especially for red and near-infrared (NIR) wavelengths.
The development of a "NIR-SiPM" is anticipated to be a breakthrough in a number of applications – first and foremost in LIDARs – as well as in medicine and health sciences, biology and physics, environmental monitoring and quantum telecommunications. To achieve efficient absorption of photons and a high PDE for the red part of the light spectrum, a thicker photosensitive layer is required due to the low absorption coefficient of silicon at these wavelengths. But, the available increase of the thickness is eliminated by the so-called “border effect”. 
The thicker the active layer of a SiPM for efficient absorption of NIR photons, the higher the losses of a sensitive volume adjacent to trenches \cite{9}. Further, the breakdown voltage increases proportionally to the depletion depth for planar technologies. 
This leads to high operating voltages in the Geiger-Mode with high temperature coefficients. Simultaneously, the lateral distance of the active area to the micro-cell edge has to be increased in order to prevent edge breakdowns at high voltages. 
%For example, the absorption efficiency of \SI{45}{\%} for 905 nm wavelenght photons requires Si thickness of 20 um, and its depletion. 
Despite these limitations, further progress in planar SiPM technology has been made with respect to the sensitivity at \SI{905}{\nano m}. For example, FBK started from \SI{11}{\%} for \SI{35}{\micro m} cells in 2017 \cite{9} and Broadcom improved this result to \SI{18}{\%} (cell size is claimed as “smallest”) in 2020 \cite{Broadcom2020}.
\IEEEpubidadjcol
\\
In contrast, there are also several photoelectric devices based on non-planar configurations.
The most relevant examples are Geiger-Mode APDs developed as predecessors or alternatives to the planar SiPMs. Spherical avalanche diodes with a radius of the pn-junction of \SI{2}{\micro m} and a breakdown voltage of \SI{50}{V} seems to be the earliest device of this type operated in a photon counting mode \cite{12}. Geiger-Mode APDs with negative feedback, also known as Metal-Resistor-Semiconductor (MRS) APDs, were designed as an array of avalanche micro-channels. A few micrometer-size n+ diffusion dots or “needle” junctions on p-Si wafer are covered by a thin resistive SiC layer as quenching resistor \cite{G,13}. The first NIR-GM-APD with negative feedback has also been developed with a similar design \cite{14}. Another kind of the microchannel GM-APD, the Micropixel APD (MAPD), is based on deep n-type micro-dots buried into a p-type epi-layer \cite{H,15}. The MAPDs demonstrated a unique pixel-density of up to $\SI{4e4}{mm^{-2}}$ with the highest geometric efficiency of almost \SI{100}{\%}. However, the MAPD design is also associated with a long pixel recovery time eliminating its dynamic range for continuous illumination and long light pulse detection.

For years, KETEK has been providing state-of-the-art SiPMs optimized for a blue range of the light spectrum, which is commonly used in medical and high energy physics applications.
To overcome the above mentioned limitations of the planar technology, we developed a new SiPM concept according to the patent application \cite{Patent_TAPD}, the "Tip Avalanche Photodiode" (TAPD).
The concept is based on utilizing the properties of tip-like electrodes to focus and enhance the electric field, to reduce the breakdown voltage and cell capacitance and to eliminate the needs in a peripheral separation of the SiPM cells (avalanche regions). 

In this article, we give an in-depth overview of the physical model of our new SiPM design and explain the working principle of the device. Further, we present the metrological characterization of our prototype samples and put them in relation to our simulations.

\section{Concept and Physical Model}
\begin{figure}[b]
\centering
\includegraphics[width=\columnwidth]{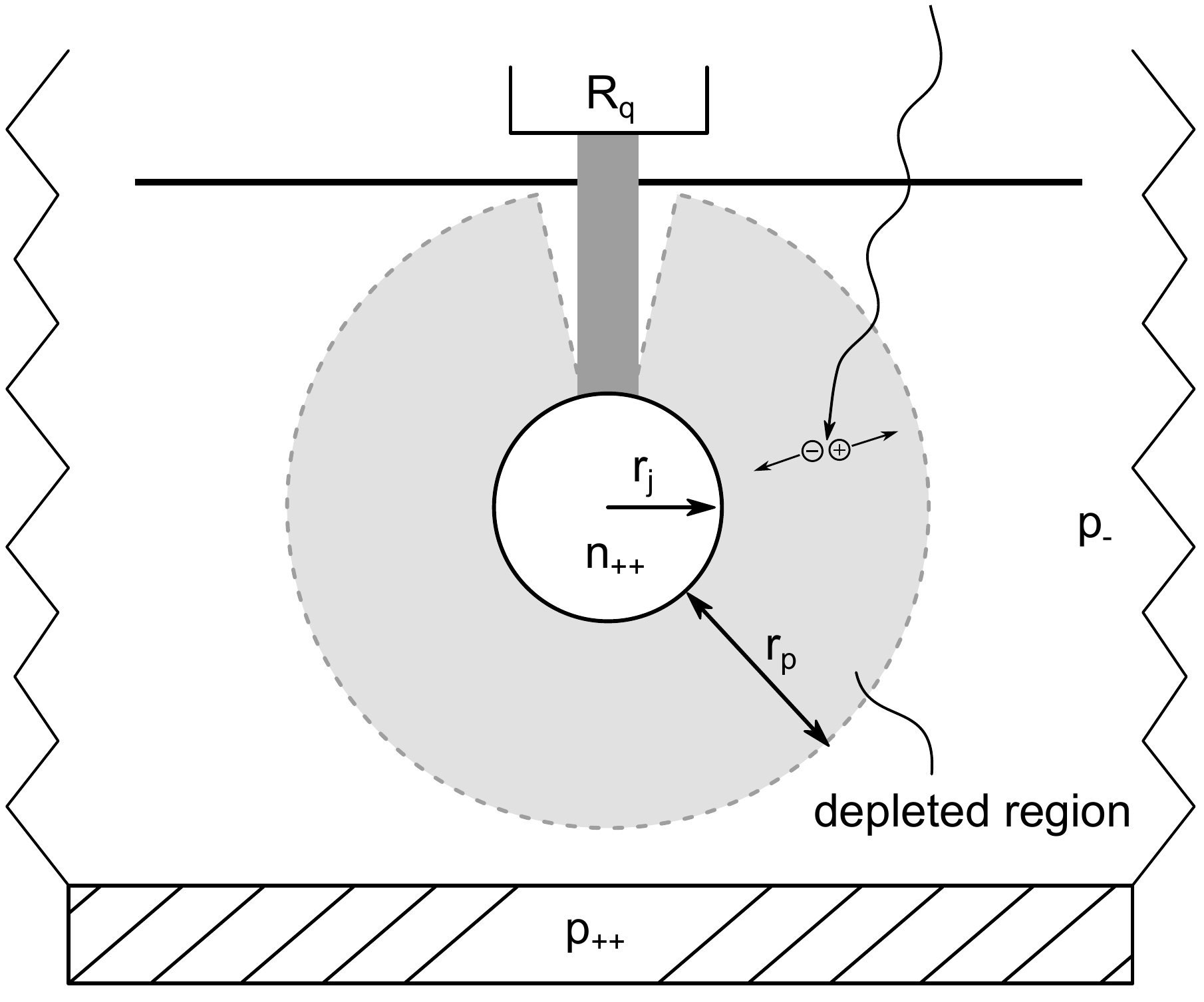}
\caption{Schematic of the spherical SiPM concept}
\label{fig:schematic_concept}
\end{figure}

The simplified cross section of the TAPD is shown in Fig.~\ref{fig:schematic_concept}.
A spherical tip which consists of high doped n-silicon is placed in a low doped epitaxial layer.
The tip is connected to the bias supply through a quenching resistor $R_q$ placed on the surface.
The pn-junction on the surface of the tip causes a depletion and therefore an electric field around the sphere.

\subsection{Analytic Description}

In this article, the basic properties of the new SiPM concept are derived using a simplified model of an n-doped sphere inside an infinite p-doped bulk. The transition from n-doped to p-doped region is first approximated as an abrupt junction where the depleted charge has a box profile (in 3D spherical shells).

\begin{figure}
\centering
\includegraphics[width=7.5cm]{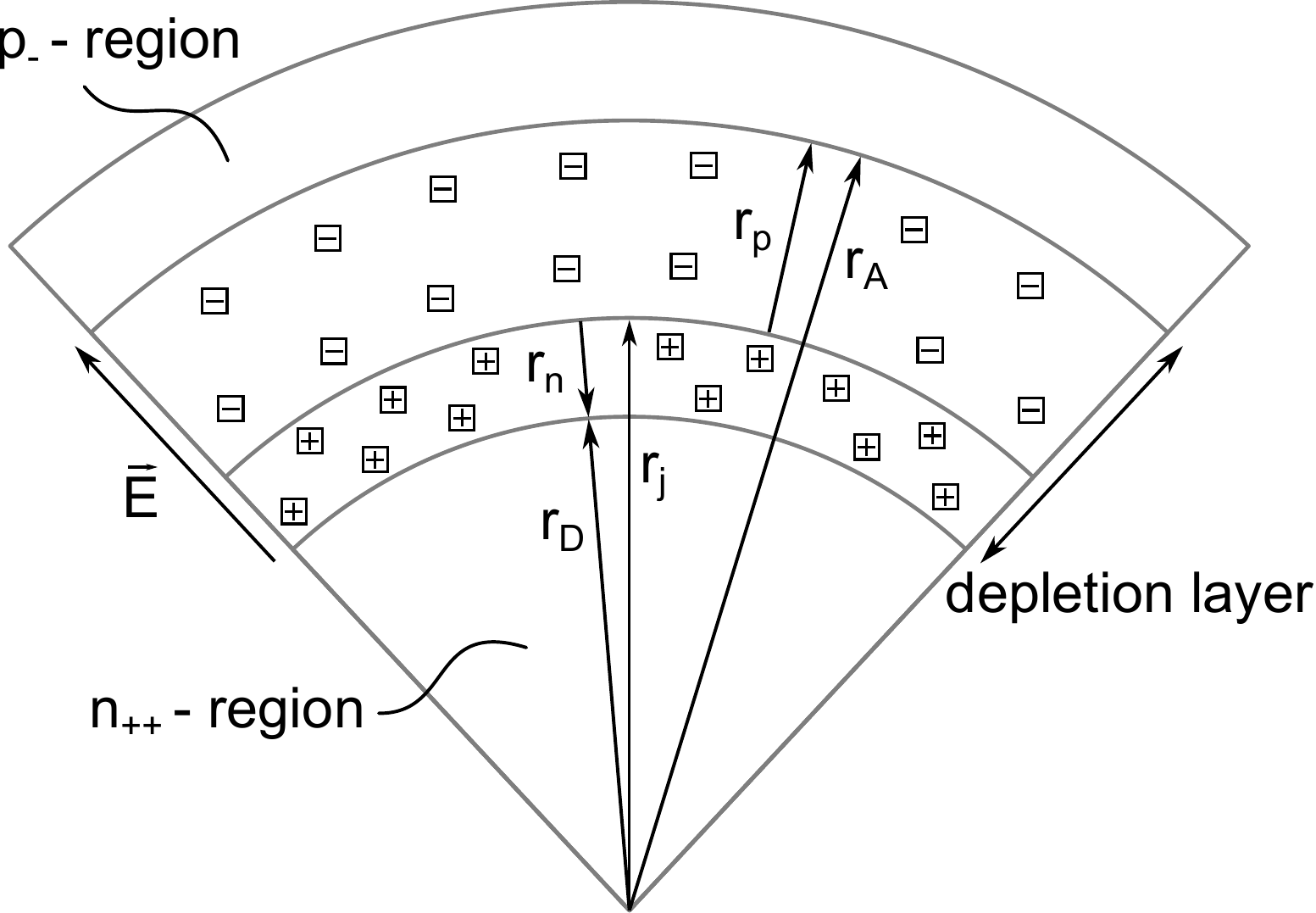}
\caption{Dimensions and depletion layers of the spherical pn-junction model.}
\label{fig:concept_model}       % Give a unique label
\end{figure}

In thermal equilibrium, the charge carrier currents of drift and diffusion cancel out and the Fermi level $E_F$ is constant throughout the junction:
\begin{equation}
\begin{split}
J_{n} = &  q \mu_n \left( n \mathcal{E} + \frac{kT}{q} \frac{dn}{dr} \right) = 0  \\
J_p = & q \mu_p \left( p \mathcal{E} + \frac{kT}{q} \frac{dp}{dr} \right) = 0
\end{split}
\label{eq:zero_net_current}
\end{equation}
The depletion approximation considers complete impurity ionization in the n- and p-region. At thermal equilibrium (subscript '$o$') the free charge carrier concentration can be simplified to $n_{no} \approx N_D$ in the n-region and $p_{po} \approx N_A $ in the p-region. 
The condition of Eq.(\ref{eq:zero_net_current}) requires a constant Fermi level and therefore a built-in voltage is present in the pn-junction. The built-in potential $\Psi_{bi}$ can be written as: 
\begin{equation}
\begin{split}
\Psi_{bi}	& =	\frac{kT}{q} \ln \left( \frac{n_{no}}{n_i}  \right) + \frac{kT}{q} \ln \left( \frac{p_{po}}{n_i}  \right) \\
			& \approx \frac{k_B T}{q} \ln \left(\frac{N_D N_A}{n_i^2}\right)
\end{split}
\end{equation}
This expression of the built-in potential is valid for planar and spherical junctions.
The electric field and the potential of the space charge region can be obtained solving the Poisson equation.
The one dimensional Poisson equation in spherical coordinates for any arbitrary charge density $\rho(r)$ is \cite{JayantBaliga1976}:
\begin{equation}
	\frac{1}{r^2} \frac{d}{d r} \left( r^2 \frac{d \Psi_i(r)}{d r} \right)  = - \frac{\rho(r)}{\varepsilon \varepsilon_0}
\label{eq:poisson}
\end{equation}

In the first case of an abrupt junction, the charge carrier densities are given by the completely ionized acceptor and donor impurities, $N_A$ and $N_D$. 
The ionized regions and the notation of the dimensional variables are shown in Fig.~\ref{fig:concept_model}. 
The analytic solution for the electric field in the n-region and p-region ($\mathcal{E}_n$ and $\mathcal{E}_p$) can be obtained from integration of Eq.(\ref{eq:poisson}) with the boundary condition of zero electric field outside the depletion region:  
\begin{align}
\begin{split}
\mathcal{E}_n(r)& = \frac{ e N_D}{3\varepsilon \varepsilon_0} \frac{\left(r^3 - r_D^3\right)}{r^2},\\
\mathcal{E}_p(r) &= \frac{ e N_D}{3\varepsilon \varepsilon_0} \frac{\left(r_j^3 - r_D^3\right)}{r^2} - \frac{e N_A}{3\varepsilon \varepsilon_0} \frac{\left(r^3 - r_j^3\right)}{r^2}
\end{split}
\label{eq:Efield_ana}
\end{align}
The electric field in the n-region is zero at the inner depletion edge $r_D$ and will increase linear if the term $r_D^3 / r^2$ is small.
In the p-region, the expression of the electric field has two terms.
The first term is dominant if $N_D \gg N_A$ and the electric field will decrease proportional to $1/r^2$. 
The second term becomes dominant if the first term is small enough with increasing radius and the electric field will equal zero at the outer depletion edge $r_A$.

\begin{figure}
\centering
\includegraphics[width=\columnwidth]{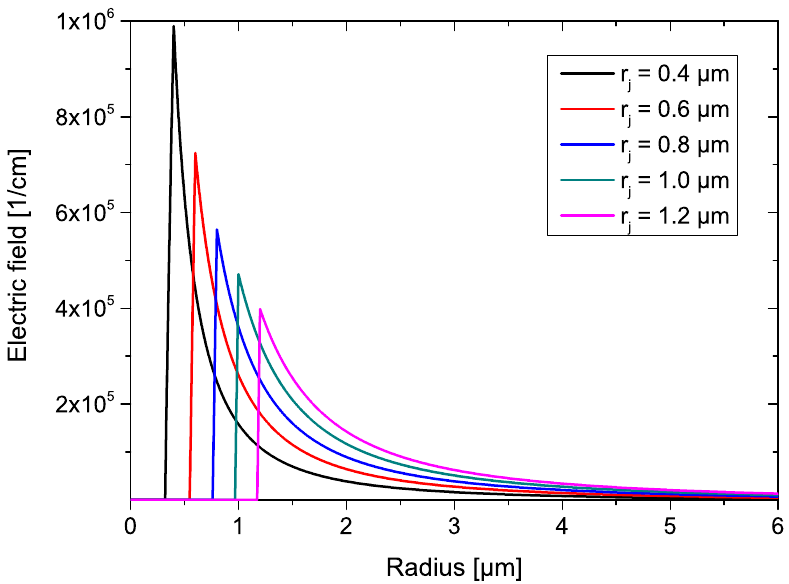}
\caption{Electric field distribution in respect with the analytic solution for different sphere radii at a depletion potential of \SI{40}{\V}.}
\label{fig:Efield_ana}
\end{figure}

In this concept, the donor concentration in the sphere is much higher than the acceptor concentration in the epitaxial layer.
A exemplary electric field distribution according to Eq.(\ref{eq:Efield_ana}) for different metallurgical junction radii is presented in Fig.~\ref{fig:Efield_ana}. Here, the donor and acceptor concentrations were arbitrarily set to $N_A = \SI{1e14}{\per \cm}$, $N_D = \SI{1e18}{\per \cm}$.
The depletion width of each sphere size was adapted to create a depletion potential of \SI{40}{\V}.

\subsection{Numerical Simulation}

The electric field in the previous section is derived using an approximation of the depletion regions as box profiles.
In a processed device, the dopant concentration will vary along the radius of the sphere.
A dopant distribution is typically created by ion implantation and thermal annealing during different process steps. 
The diffusion of impurities is thermally activated and occurs in direction of the concentration gradient.
In the case of the TAPD, especially the donor impurities of the tip will start to diffuse into the p-region.
Three different structure sizes were processed and tested. The nominal junction radii of these devices are \SI{0.6}{\micro \m}, \SI{0.8}{\micro \m} and \SI{1.0}{\micro \m}.
The measured dopant profile of these devices is used for the following simulations.

The complete impurity ionization (as assumed in Eq.(\ref{eq:Efield_ana})) is prevented by diffusion of free charge carriers into the depleted regions \cite{Sze2007}.
While Eq.(\ref{eq:zero_net_current}) deals with the steady-state, the continuity equations describe the net current flowing in and out of a region of interest. Here, the generation and recombination in the depletion region is neglected. The divergence operator has to be adapted for one dimensional spherical coordinates. The simplified continuity equations and the differential equation of the electric field from the system of coupled equation which has to be solved:
\begin{equation}
\frac{\partial n}{\partial t}	= \frac{1}{r^2} \frac{\partial}{\partial r} r^2  \left( \mu_n \mathcal{E}	n + D_n \frac{\partial n}{\partial r} \right)
\end{equation}
\begin{equation}
\frac{\partial p}{\partial t} = \frac{1}{r^2} \frac{\partial }{\partial r} r^2  \left( - \mu_p \mathcal{E} p + D_p \frac{\partial p}{\partial r} \right)
\end{equation}
\begin{equation}
\label{eq:depl}
	\frac{1}{r^2} \frac{d}{d r} r^2 \mathcal{E}(r)   = \frac{q}{\varepsilon_s} \left( p(r) - n(r) + D(r) \right)
\end{equation}
where $\mu_n$, $\mu_p$ are the electron and hole mobility and $D_n$, $D_p$ are the electron and hole diffusivities.
The spatial dependent doping profile $D(r)$ defines the p and n-region as:
\begin{equation}
D(r) = 
\begin{cases}
r \leq r_j; &N_D(r) \\
r \geq r_j;  &-N_A(r)
\end{cases}
\end{equation}

The presented governing system of equations is considered in a spherical domain.
The boundary conditions in the center and on the spherical shell of the region of interest were of the Dirichlet type. 
We used the finite difference method with a centred difference stencil for the spatial numerical approximation. 
The approximation in time of the continuity equations was done using the Crank Nicolson method \cite{crank_nicolson_1947}.

\begin{figure}
\centering
\includegraphics[width=\columnwidth]{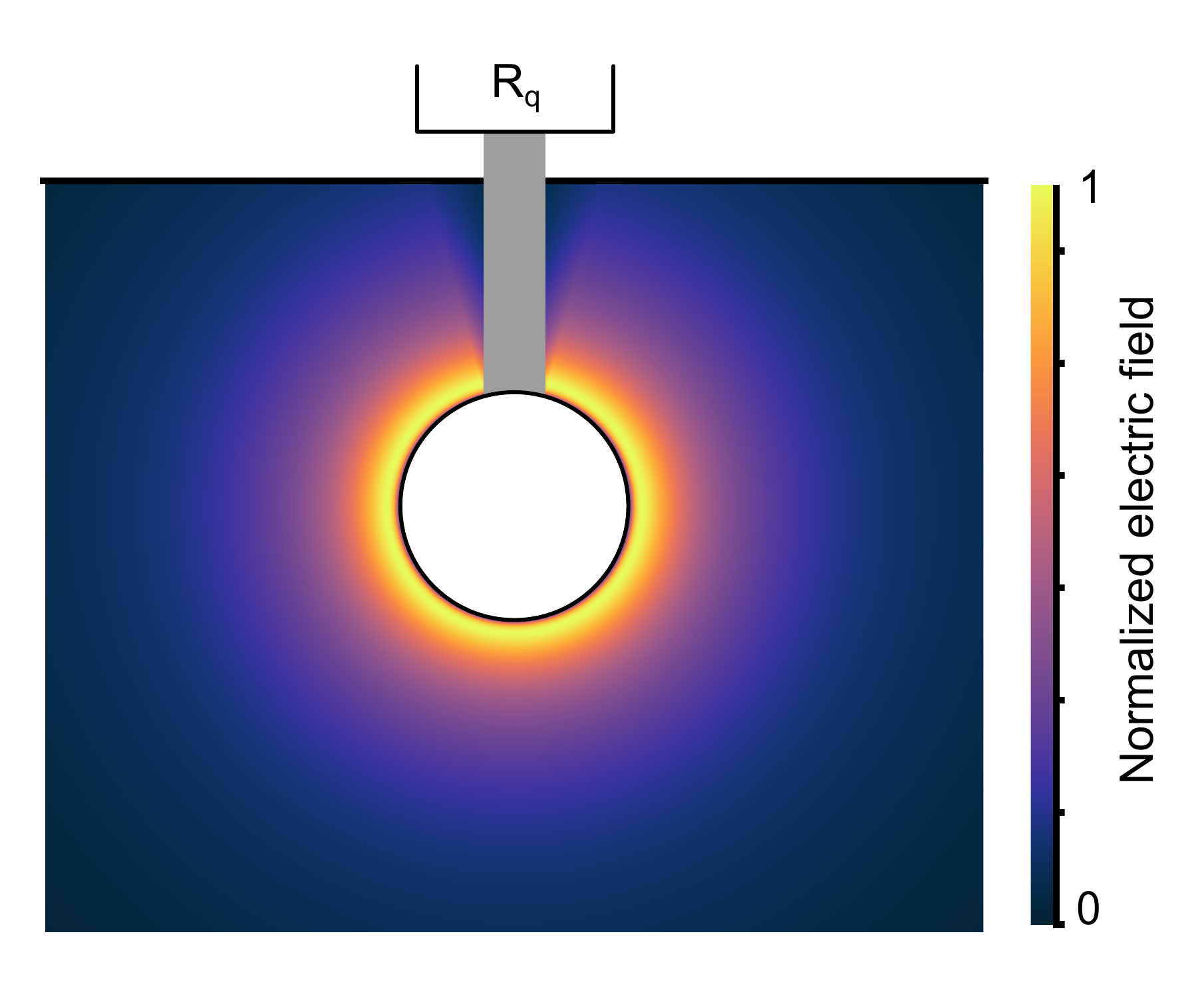}
\caption{Schematic 2D electric field distribution due to the spherical depletion around the tip.}
\label{fig:2D-field}
\end{figure}

\begin{figure}[b]
\centering
\includegraphics[width=\columnwidth]{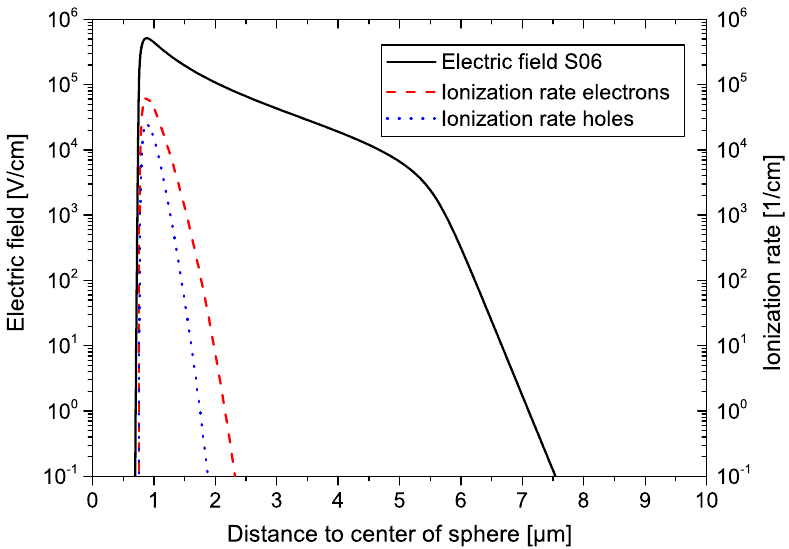} 
\caption{Spatial distribution of the electric field and the ionization rates of electrons and holes for the smallest structure S06.}
\label{fig:plot_Efield}      
\end{figure}

\begin{figure*}[bp]
\centering
\includegraphics[width=14cm]{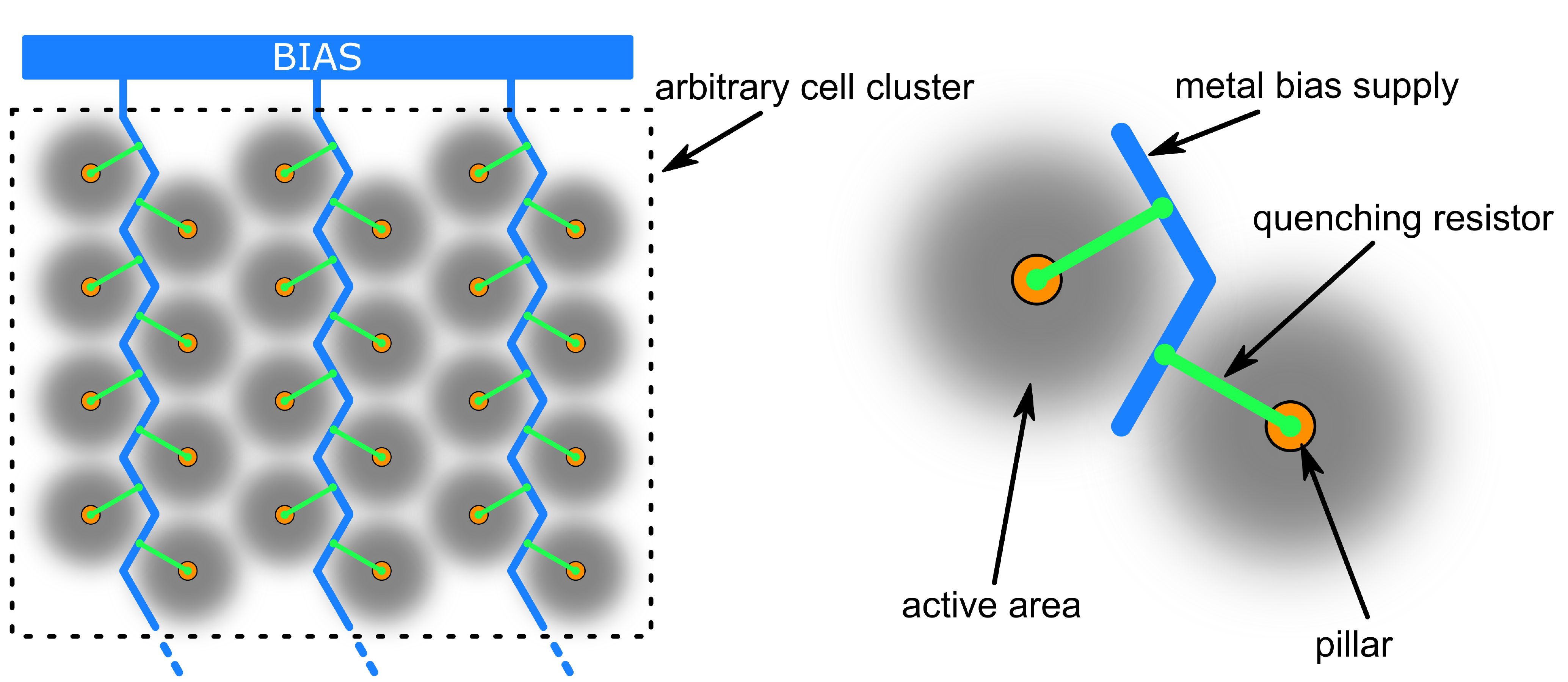}
\caption{High density layout and bias supply of a TAPD array.}
\label{fig:layout}
\end{figure*}

In Fig.~\ref{fig:2D-field}, a schematic 2D distribution of the normalized electric field around the tip is illustrated. The highest field strength is located close to tip surface and decreases with increasing distance to the junction.
The electric field is present up to the passivated surface and a spherical active volume is available for electron attraction in direction of the n-region.
However, impact ionization multiplication due to a high electric field occurs just close to tip surface.
The solution of the numerical approximation at the breakdown voltage (here \SI{43.4}{V}) is presented in Fig.~\ref{fig:plot_Efield} for a nominal junction radius of \SI{0.6}{\micro \m}.
The ionization rates $\alpha_n$ and $\alpha_p$ for electrons and holes depend on the electric field:
\begin{equation}
\alpha(\mathcal{E}) = \alpha_\infty \exp\left(\frac{-b}{\mathcal{E}} \right)
\end{equation}
with $\alpha_\infty$ and $b$ at room temperature taken from \cite{VanOverstraeten1970} for electrons and holes, respectively.
In order to estimate the breakdown voltage of different sphere sizes, the multiplication of charge carriers due to impact ionization has to be evaluated using the spatial dependent ionization rates at different bias voltages.
A breakdown occurs if the impact ionization multiplication becomes infinite which is equivalent to the condition that the ionization integral \cite{McINTYRE1966} equals one:
\begin{equation}
\int_{r_D}^{r_A} \alpha_n \exp\left[  - \int_{r}^{r_A}  \left( \alpha_n - \alpha_p  \right) dr' \right] dr = 1
\end{equation}
The results of this evaluation are presented in Tab.~\ref{tab:structures}. The breakdown voltage increases with the size of the tip due to a lower electric field at the same bias voltage.
\begin{table}[h]
\caption{Tested device structures and nominal radii of the metallurgical junction.}
\label{tab:structures}       % Give a unique label
\begin{tabular}{lll}
\hline\noalign{\smallskip}
Structure Name &  Nominal Radius ($r_j$) & Breakdown Voltage \\
\noalign{\smallskip}\hline\noalign{\smallskip}
S06 & \SI{0.6}{\micro \m} & \SI{43.4}{V} \\
S08 & \SI{0.8}{\micro \m} & \SI{50.7}{V} \\
S10 & \SI{1.0}{\micro \m} & \SI{53.9}{V} \\
\noalign{\smallskip}\hline
\end{tabular}
\end{table}
The range of the electric field directly affects the photon detection capability regarding light with increasing wavelength.
The absorption of a photon inside of a material at a certain depth $x$ can be described by the Beer-Lambert law.
The probability for the generation of charge carriers at a certain depth inside the epitaxial layer decreases exponentially with increasing distance to the surface.
Additionally, in the range of visible light, the absorption coefficient decreases with increasing wavelength.
A SiPM with a high charge collection efficiency aims to absorb as much of the incoming photon flux as possible. 
Consequently, the electric field has to be present as deep as possible.
Photoelectrons generated at a distance to the tip are first accelerated due the drift field in direction of the multiplication region (high electric field) and a Geiger discharge is triggered.

The simulated range of the electric field into the epitaxial layer for the three structure sizes at different bias voltages is shown in Fig.~\ref{fig:depletion_width}.
All structures reach at least \SI{8}{\micro \m} at their respective breakdown voltage.
The maximal active volume is reached placing the center of the sphere at a depth of the maximal depletion range. In this configuration, photoelectrons which are generated close to the surface and up to the maximal active depth are detected. This advantage of the new concept allows for a high photon detection efficiency in a wide spectral range. Theoretically, it is possible to reach a total active depth of \SI{20}{\micro m} with the current technology (see Fig.~\ref{fig:depletion_width}) and consequently 
a photon detection efficiency above \SI{30}{\%} at \SI{905}{nm}.
Our prototypes were processed in an epitaxial layer of \SI{12}{\micro m}, which limits the maximal depleted volume for all structure types.

\subsection{Cell Placement}

The operation of a SiPM above the breakdown voltage in Geiger-Mode requires a serial connected quenching resistor to keep the device in a quasi-stable state. 
The potential of the epitaxial layer during operation is set to ground while the tip is biased through the resistor.
The spherical shape of the active volume allows a high density placement of multiple single cells in an array.
In the top view, the cells are placed in a hexagonal lattice achieving a theoretical packing density of $\eta \approx \SI{90.7}{\percent}$.
The bias voltage is supplied through a metal grid connected to the quenching resistor of each cell.
The schematic layout is shown in Fig.~\ref{fig:layout}.
The aim of the layout is to cover the smallest possible area with metal. 
The metal on the surface blocks visible light therefore reducing the active area of the SiPM for light detection.
The presented placement offers the advantage to use just one metal line for two rows of SiPM cells.

The relation of uncovered area to total area is called geometric efficiency (see also Sec.~\ref{sec:PDE}).
Compared to planar devices, the new concept offers the advantage of a frameless layout. The TAPD cells are biased through the center and can be placed as close as possible next to each other without losing active area.

A typical parameter for SiPMs is the cell pitch which equals the cell size for planar devices. We defined the cell pitch for the TAPD as the distance of pillars (see Fig.~\ref{fig:layout}).
A small cell pitch offers the advantage of a higher dynamic range \cite{Vinogradov2011}.
The absorption of a photoelectron in the active volume triggers a Geiger-discharge of a single cell. During the discharge and the following recharge of the cell, the cell is partially inactive for incoming light \cite{Oide2007,Renker2009}.
Consequently, the SiPM array shown in Fig.~\ref{fig:layout} could only detect a maximum of 24 photons during a fast light pulse (e.g. shorter than recharge time).

\begin{figure}[t]
\centering
\includegraphics[width=\columnwidth]{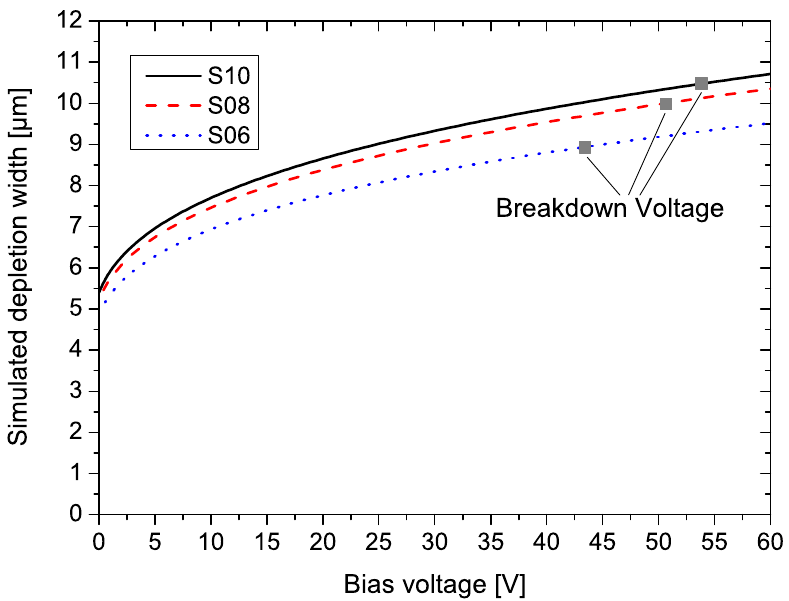}
\caption{Range of the electric field for the three structure sizes.}
\label{fig:depletion_width}
\end{figure}

The aim of a high dynamic range and a high geometric efficiency is in general contradictory.
Regarding the results of Fig.~\ref{fig:depletion_width}, we chose a cell pitch of \SI{15}{\micro \m} for the processed devices.
The active areas of two neighbouring cells for the smallest structure S06 are just in contact.
The geometric efficiency of our device is only determined by the opaque materials like metal lines and semi-transparent materials like the quenching resistors, which are located on top of the active area. 
For the prototypes with a \SI{15}{\micro m} pitch, the nominal geometric efficiency of \SI{83}{\%} was realized. The value was calculated from the layout, accounting metal lines and quenching resistors as opaque.

\subsection{Depletion Layer Capacitance}

The depletion layer around the tip creates a certain amount of charge on each side of the junction whereas the total charge is zero.
The incremental charge $dQ_D$ of one side of the junction upon an incremental change of the bias voltage defines the depletion layer capacitance $C_D = dQ_D / dV$.
The spatial depletion of the respective structures was obtained with the solution of the continuity equations (see Eq.(\ref{eq:depl})).
The simulated depletion layer capacitance dependent on the bias voltage is presented in Fig.~\ref{fig:depletion_cap}.
All three structures show a similar curve progression while the increased tip size leads to a higher capacitance.

The simulated depletion layer capacitance is part of the total cell capacitance $C_{\mathrm{cell}}$, which includes additionally the parasitic capacitance due to the pillar and the connection to the quenching resistor.
The gain of a single cell is proportional to the cell capacitance and can be expressed as:  
\begin{equation}
G = \frac{Q}{e} = \frac{\Delta V \cdot C_{\mathrm{cell}} }{e}
\end{equation}
where $\Delta V$ stands for the excess bias voltage.
The main goal for the presented new SiPM design was to a achieve a high photon detection efficiency over a wide spectral range.
The recovery time (recharge time) should be short to provide a fast device with a high dynamic range. Consequently, a low cell capacitance is beneficial.
In respect with the simulations of this section, we chose the smallest device S06 as the most promising structure.

\section{Metrological Characterization}
\subsection{Single electron response}
\label{sec:SER}
\begin{figure}[h!]
\centering
\resizebox{\columnwidth}{!}{\includegraphics{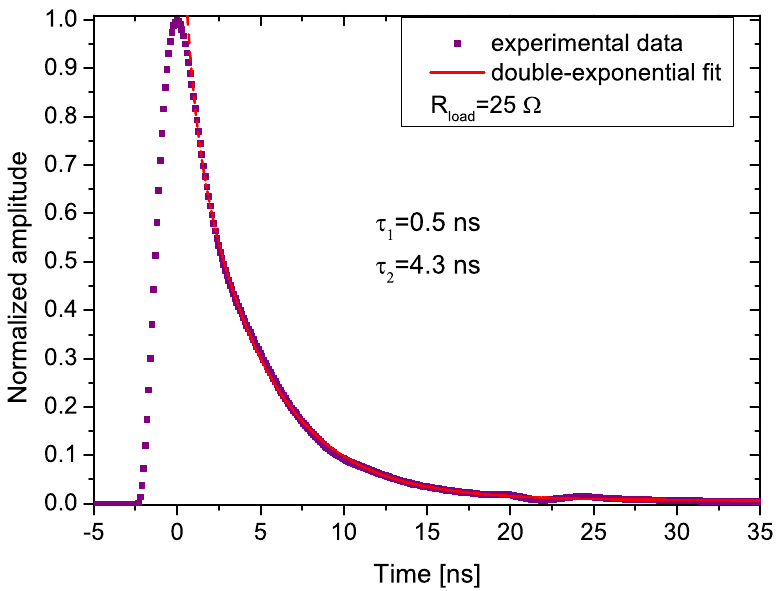}}
\caption{Normalized pulseshape of the S06 structure at an excess bias voltage of \SI{4}{V}. }
\label{fig:SER}       % Give a unique label
\end{figure}

To achieve a high photon count rate and a strong ambient light immunity, a fast recovery of the micro-cells is required. In Fig.~\ref{fig:SER}, the single electron response (SER) of the 3D-SiPM is shown at $\Delta V=4\text{\,V}$.
It was measured as the voltage drop across a \SI{25}{\Omega} load resistor. 
%The rise-time of the pulse is about \SI{1.3}{\nano s}. 
The decay part of the pulse consists of two exponential components with time constants of $\tau_{1}\approx \SI{0.5}{\nano s}$ and $\tau_{2}\approx \SI{4.3}{ns}$.
After approximately \SI{9.5}{\nano s}, the micro-cells are recovered to \SI{90}{\%} of their maximum charge. Using the double-light-pulse method, as proposed in \cite{Popova2012}, comparable results were achieved.
This confirms that the SER shape reflects the true recovery process in this case. With this result, the TAPD has the fastest recovery time with respect to state-of-the-art planar SiPMs with enhanced red-sensitivity \cite{Broadcom2020,RB_Series2020}. 

\begin{figure}
\centering
\resizebox{\columnwidth}{!}{\includegraphics{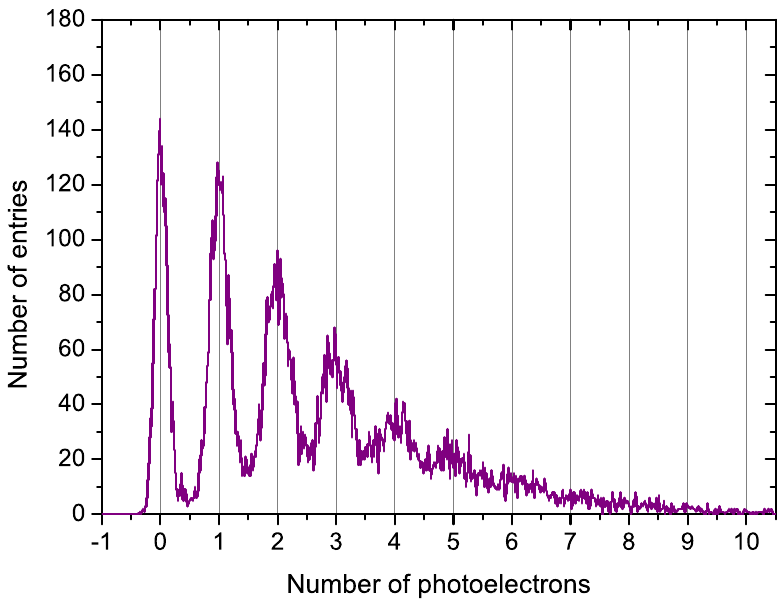}}
\caption{Single photoelectron charge spectrum of the S06 structure at an excess bias voltage of \SI{4}{V}.}
\label{fig:SPE_Spectrum}       % Give a unique label
\end{figure}

Fig.~\ref{fig:SPE_Spectrum} provides an example of a single photoelectron charge spectrum. For this spectrum, we used a pulsed laser illumination with a pulse width of \SI{70}{ps}. The acquisition was synchronous with the light pulses. The charge was integrated within a time-window of \SI{10}{\nano s}. The peaks up to 4 photoelectrons are well separated, which makes precise single photon counting possible.
%

%------------------------------------------------------------------------------------------------------------------
%------------------------------------------------------------------------------------------------------------------
\subsection{Photon detection efficiency}
\label{sec:PDE}
The photon detection efficiency (PDE) describes the capability of the sensor to detect light as the ratio of the average number of incident and the average number of detected photons. In Eq.(\ref{eq:PDE_Def}), the PDE is described as a product of three quantities:

\begin{equation}
\label{eq:PDE_Def}
PDE=\varepsilon \cdot QE \cdot P_{trigg}
\end{equation}

(i) The geometric efficiency $\varepsilon$ describes the fraction of the SiPM area, which is able to detect photons. The area of the SiPM which is not sensitive to light is mainly due to the metal lines for signal readout, the quenching resistors, guard rings for electric field attenuation towards the micro-cell edges and trenches for the suppression of optical crosstalk.

(ii) The quantum efficiency $QE$ describes the efficiency to collect a fraction of charge carriers that a photon generated within the active volume of a micro-cell.
%The quantum efficiency $QE$ describes the probability that a photon is absorbed within the active volume of a micro-cell and generates an e-h pair. 
To maximize the PDE for blue light, the depletion region has to be extended as close to the surface as possible. Lower energetic photons are also absorbed at larger depths. To reach an enhanced detection of red and near-infrared light, the active region has to reach deeper inside the silicon. With our devices, we reach a spherical depletion volumes with radii of about \SI{8}{\micro m} to \SI{9}{\micro m} (see Fig.~\ref{fig:depletion_width}).

\begin{figure}
\centering
\resizebox{\columnwidth}{!}{\includegraphics{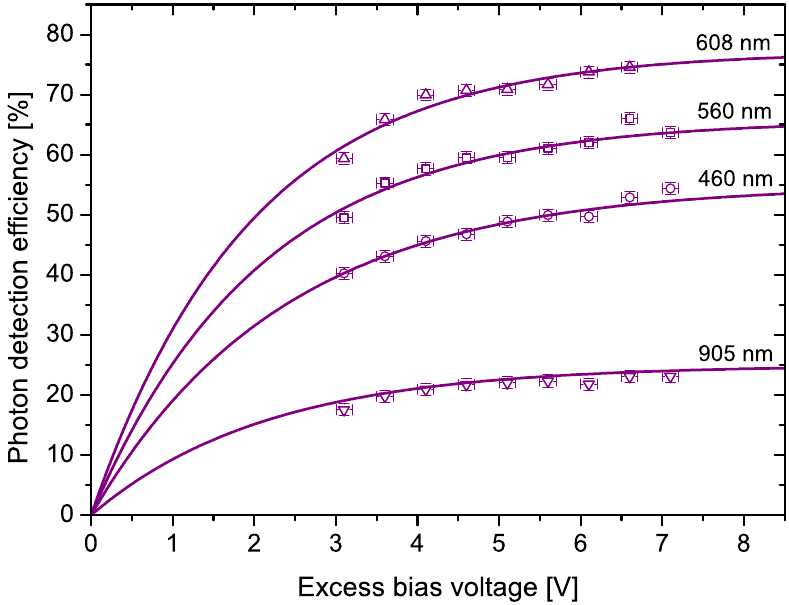}}
\caption{Photon detection efficiency vs. the excess bias voltage for the S06 structure. The lines are drawn as eye guides.}
\label{fig:PDE_vs_EBV}       % Give a unique label
\end{figure}

(iii) The avalanche triggering probability $P_{trigg}$ describes the probability that a generated e-h pair will successfully initiate an avalanche breakdown by impact ionization. 
   
In this work, the PDE was measured by using the continuous low-level light method, as reported in \cite{Piemonte2012,Vachon2018,Vinogradov2016,Engelmann2018}. The incident photon rate was determined by a calibrated reference PIN-diode \cite{Thorlabs2019}. Both, the SiPM and the reference diode were homogeneously illuminated trough a lens. The homogeneous part of the light spot was significantly larger than the active area of the photosensors. The photon rate was determined as the difference of the SiPM pulse count rate during illumination and at dark conditions. The count rates were measured according to the method described in Sec.~\ref{sec:DCR_PCP}.
In Fig.~\ref{fig:PDE_vs_EBV}, the PDE is shown as a function of the excess bias voltage for several wavelengths from \SI{460}{\nano m} to \SI{905}{\nano m}. In this spectral range, the PDE reaches \SI{90}{\%} of its saturation value at excess bias voltages between \SI{4}{V} and \SI{5}{V}. This is comparable to state-of-the-art planar SiPMs which are optimized for the detection of blue light and hence have smaller depletion volumes \cite{Otte2017}.

\begin{figure}[b]
\centering
\resizebox{\columnwidth}{!}{\includegraphics{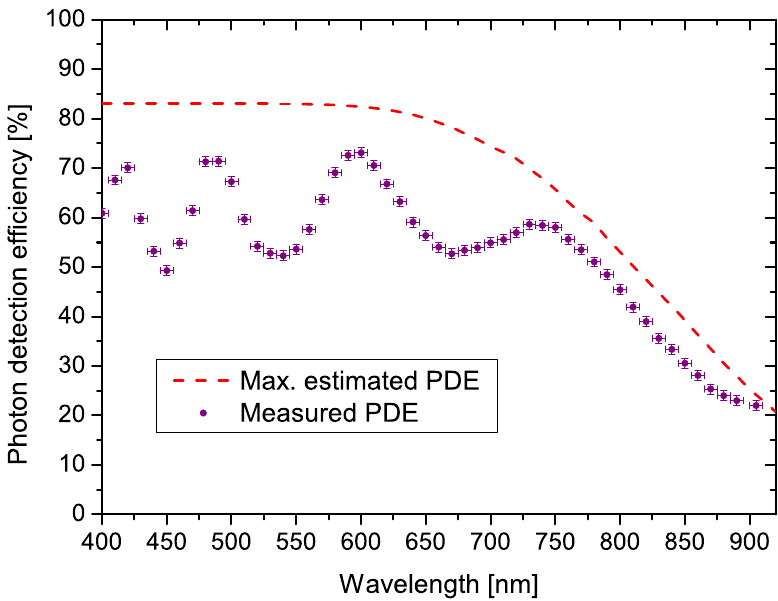}}
%\resizebox{0.45\textwidth}{!}{\includegraphics{PDE_vs_WL.pdf}}
\caption{Photon detection efficiency vs. wavelength for the S06 structure at an excess bias voltage of \SI{5}{V}.}
\label{fig:PDE_vs_WL}       % Give a unique label
\end{figure}

In Fig.~\ref{fig:PDE_vs_WL}, the PDE is shown at different wavelengths. This result was obtained by converting the spectral response measurement with a monochromator into an absolute PDE measurement as described in \cite{Otte2017}.
The TAPD demonstrates the highest peak PDE value of \SI{73}{\%} at \SI{600}{\nano m} with respect to state of the art devices with a \SI{15}{\micro m} pitch size \cite{Gola2019,Yamamoto2018}. 
Additionally, the PDE curve does not show the typical fast decrease with increasing wavelengths and remains above a value of \SI{45}{\%} up to a wavelength of $800\text{\,nm}$. The measured PDE is in good agreement with the expected values when assuming a Geiger-efficiency of \SI{90}{\%} (see Eq.(\ref{eq:PDE_Def}) and dashed curve in Fig.~\ref{fig:PDE_vs_WL}).
For wavelengths in the NIR-regime, we expect the geometric efficiency to increase approximately to \SI{90}{\%} due to the decreasing absorption by the semi-transparent quenching resistors.
The oscillations in the PDE curve are caused by destructive interference due to the protective $Si/SiO_2$ stack on top of the entrance window of the prototype devices.
The results of the applied method are in agreement with the well-known method based on pulsed laser illumination \cite{Eckert2010,Otte2017}.
The applied method offers the possibility to directly measure the absolute PDE for a larger number of wavelengths due to the easy access to LEDs of different wavelengths.

%------------------------------------------------------------------------------------------------------------------
%------------------------------------------------------------------------------------------------------------------
\subsection{Dark count rate and delayed correlated pulses}
\label{sec:DCR_PCP}

\begin{figure}[b]
\centering
\resizebox{\columnwidth}{!}{\includegraphics{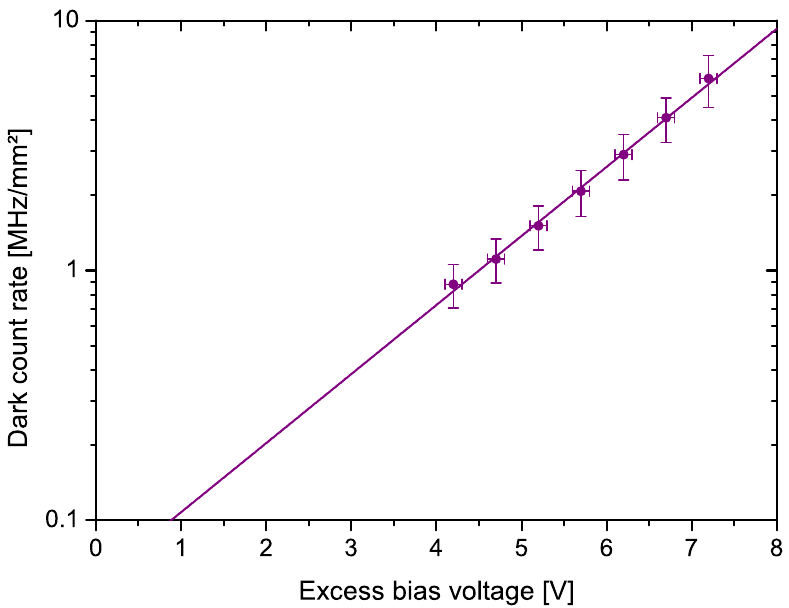}}
\caption{Dark count rate vs. the excess bias voltage for the S06 structure. The error bars represent the standard deviation of 5 samples. The line is drawn as an eye guide.}
\label{fig:DCR_vs_EBV}       % Give a unique label
\end{figure}

In this work, the dark count rate and the probability of delayed correlated pulses is determined by using the method proposed in \cite{Vinogradov2016}. The method is based on the analysis of the complementary cumulative distribution function (CCDF) of pulses subsequent to a primary dark pulse. Contrary to the pulse counting approach, the applied method provides the benefit that delayed correlated pulses which exceed the detection threshold do not contribute to the dark count rate of the device \cite{Engelmann2018}. This advantage is of special significance for SiPMs with a fast recovery time, like the TAPD, where delayed correlated pulses reach the maximum charge after a few ns.
The CCDF-method is also suited to measure the count rate of a continuous Poissonian photon output from a light source. In this way the photon detection efficiency can be measured (cf. Sec.~\ref{sec:PDE}).

As delayed correlated pulses, we understand pulses which may be caused by two kinds of effects: 
\\
The first one is afterpulsing. Here, trapping centres present as energy states within the bandgap may capture electrons or holes from the conduction or valence band and re-emit them after a certain delay-time $\Delta t$ into the same band. If the trapping centre is located in the active region of a micro-cell, the re-emitted charge carrier has a finite probability to trigger a subsequent avalanche in the same micro-cell. The delay-times depend on the respective trap type and may vary by many orders of magnitude.
\\
The second effect is the so called delayed optical crosstalk.
Here, one or more photons which are emitted during the avalanche breakdown are absorbed outside the high field region. The generated minority charge carriers then diffuse towards the active region and are able to trigger consecutive breakdowns of the original or a neighbouring micro-cell. This process is significantly slower compared to the prompt optical crosstalk.

The applied method is only valid under the assumption that the time constants of the delayed correlated effects are smaller with respect to the reciprocal of the dark count rate. In \cite{Engelmann_PhD_2018}, the applied measurement procedure is described in detail.

\begin{figure}[b]
\centering
\resizebox{\columnwidth}{!}{\includegraphics{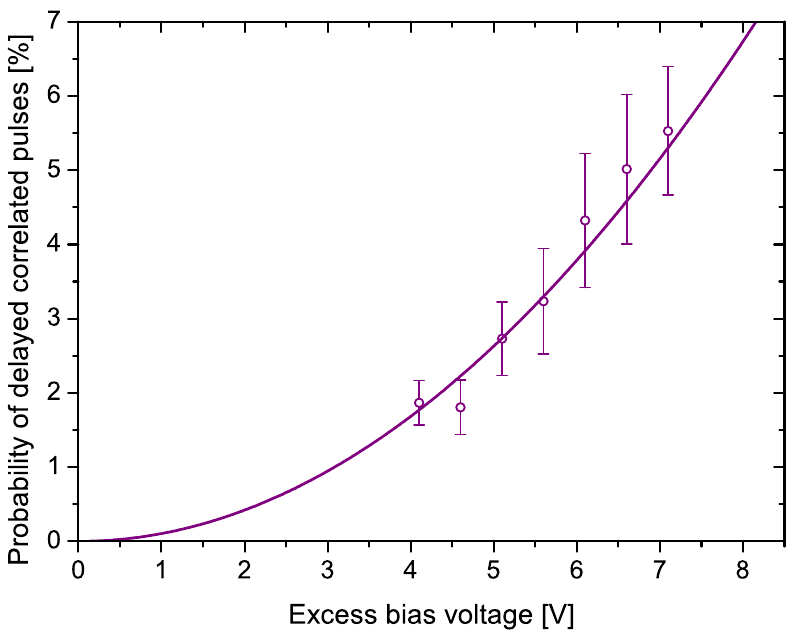}}
\caption{Probability of delayed correlated pulses vs. the excess bias voltage for the S06 structure. The error bars represent the standard deviation of 5 samples.}
\label{fig:PCP_T0P6}       % Give a unique label
\end{figure}

In Fig.~\ref{fig:DCR_vs_EBV}, the dark count rate is shown as a function of the excess bias voltage.
At recommended excess bias voltages between $\Delta V=\SI{4}{V}$ and $\Delta V=\SI{5}{V}$, the prototypes show a DCR between $\SI{700}{kHz/mm^2}$ and $\SI{1.3}{MHz/mm^2}$. In comparison, KETEK's planar \SI{15}{\micro m} pitch SiPM shows a DCR of typically $125\text{\,kHz}/\text{mm}^2$ at $\Delta V=\SI{5}{V}$ \cite{KETEK2019}.

This result matches our expectations, since the active volume of the TAPD is about a factor 10 larger with respect to the planar structures. The state of the art red-sensitive SiPMs from other manufacturers show dark count rates between $\SI{600}{kHz/mm^2}$ \cite{Broadcom2020} and $\SI{3.5}{MHz/mm^2}$ \cite{RB_Series2020} at the recommended operation voltages.
In Fig.~\ref{fig:PCP_T0P6}, the probability of delayed correlated pulses is plotted as a function of the excess bias voltage. It is below \SI{3}{\%} at excess bias voltages up to $\Delta V=\SI{5}{V}$.

%------------------------------------------------------------------------------------------------------------------
%------------------------------------------------------------------------------------------------------------------
\subsection{Prompt optical crosstalk}
During an avalanche breakdown, optical photons are generated by a variety of processes. These photons are able to propagate to neighbouring micro-cells and initiate further avalanche breakdowns. The propagation may occur by a direct path or by several reflections at the top and bottom side of the device. In either case, the time difference between the first and the consecutive pulse is not sufficient for a distinction of the two pulses. 
For this reason, only one pulse with an amplitude of multiple p.e. (photoelectron equivalent) is registered. This effect is called "prompt optical crosstalk" (CT). The prompt optical crosstalk probability scales with the number of generated photons during an avalanche breakdown, the geometric cross-section for the interaction between two micro-cells and the avalanche triggering probability.
The prompt optical crosstalk probability $P_{CT}$ is estimated as shown in Eq.(\ref{eq:P_CT_estimate}). Here, $DCR_{n}$ is the dark count rate measured by the pulse counting method with a discriminator threshold of $n\text{ p.e.}$ \cite{Renker2009,Eckert2010,ONeill2015,Otte2017}:
\begin{equation}
\label{eq:P_CT_estimate}
P_{CT}\approx \frac{DCR_{1.5}}{DCR_{0.5}}
\end{equation}
At typical excess bias voltages between \SI{4}{V} and \SI{5}{V}, we measure crosstalk probabilities between \SI{27}{\%} and \SI{35}{\%}. For comparison: The reported values from other state-of-the-art devices with a red-enhanced sensitivity are between \SI{20}{\%} and \SI{43}{\%} \cite{Broadcom2020,RB_Series2020}.

\begin{figure}[h!]
\centering
\resizebox{\columnwidth}{!}{\includegraphics{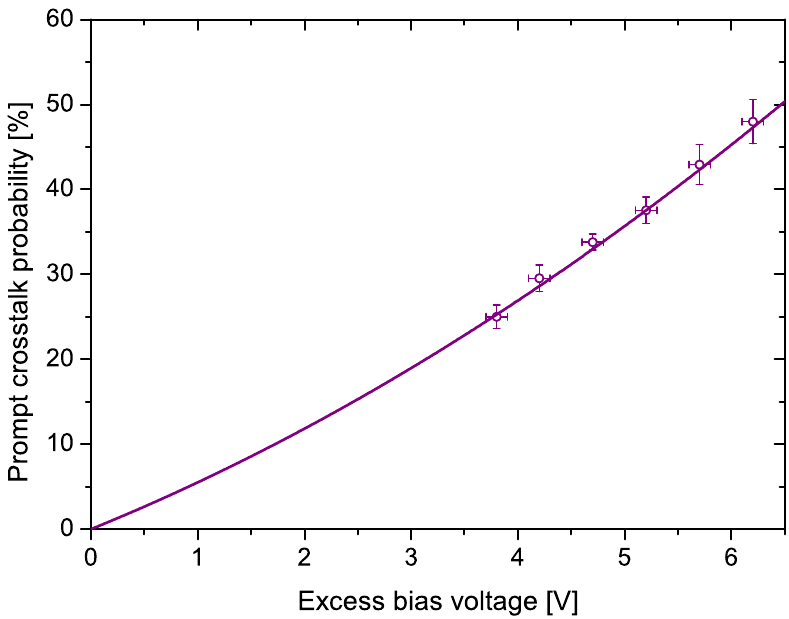}}
\caption{Prompt optical crosstalk probability of the S06 structure.}
\label{CTP_vs_EBV}       % Give a unique label
\end{figure}

%------------------------------------------------------------------------------------------------------------------
%------------------------------------------------------------------------------------------------------------------
\FloatBarrier
\subsection{Gain}
\label{sec:Gain}

\begin{figure}
\centering
\resizebox{\columnwidth}{!}{\includegraphics{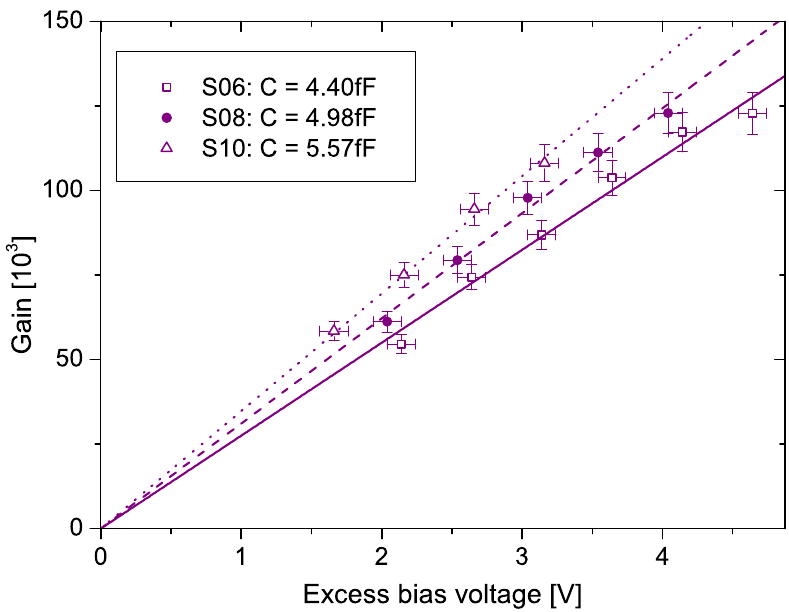}}
\caption{Gain vs. the excess bias voltage for the S06, S08 and the S10 structure.}
\label{fig:Gain_vs_EBV}       % Give a unique label
\end{figure}

\begin{figure}
\centering
\includegraphics[width=\columnwidth]{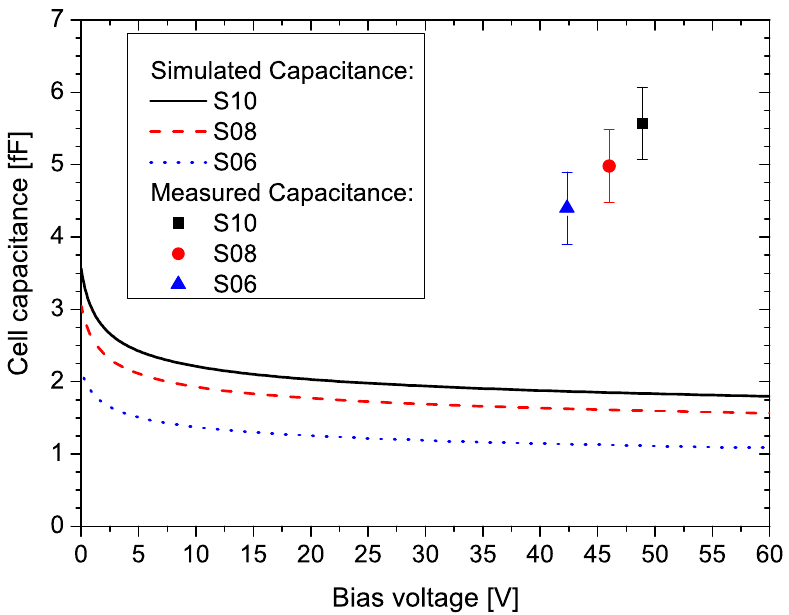}
\caption{Simulated depletion layer capacitance and measured total cell capacitance (plotted at $V_{bd}$).}
\label{fig:depletion_cap}
\end{figure}

The fast recovery time, described in Sec.~\ref{sec:SER}, is realized by a strongly reduced micro-cell capacity. As a consequence, the intrinsic gain is about a factor of 20 lower with respect to the KETEK's planar SiPM products \cite{KETEK2019}. In Fig.~\ref{fig:Gain_vs_EBV}, the absolute gain is shown as a function of the excess bias voltage for two different structures.
The higher gain of the S10 structure can be attributed to the larger ball radius and the larger contribution from the parasitic capacitance of the quenching resistor.
To measure the gain, we applied a combination of the single photoelectron charge spectrum and the dark current \cite{Pagano2012,Engelmann_PhD_2018}. 
In this procedure, the single photoelectron spectrum was recorded in dark environment with a trigger threshold set to \SI{0.5}{p.e.}.
To receive the probability density function (PDF) of the primary and secondary events, the spectrum was normalized to the total number of detected events.
The expected number of firing micro-cells per initial photoelectron is described by the mean value of the PDF which is equivalent to the excess charge factor (ECF) \cite{Klanner2019}. The dark count rate was determined as described in Sec.~\ref{sec:DCR_PCP}. In this case, the contribution from afterpulses is neglected. Since the investigated devices have a low afterpulsing probability (see Fig.~\ref{fig:PCP_T0P6}), this approach is reasonable. The gain $G$ is then determined by using Eq.(\ref{eq:Gain}). $I_{dark}$ is the dark current at the respective operation voltage:
\begin{equation}
\label{eq:Gain}
G(V)=\frac{I_{dark}(V)}{q\cdot DCR(V) \cdot ECF(V)}
\end{equation}
From the slopes of the linear fits in Fig.~\ref{fig:Gain_vs_EBV}, we extract the micro-cell capacitances. The experimentally determined micro-cell capacities differ from the simulated ones by about \SI{3.3}{\femto F} to \SI{3.8}{\femto F} (see Fig.~\ref{fig:depletion_cap}). We attribute this discrepancy to the parasitic capacitance of the quenching resistor and variation of the real geometry from the simulated spherical one. Here, we would like to point out that we expect the parasitic capacitance to be larger than the micro-cell capacitance and hence strongly contribute to the charge output of the TAPD.
%------------------------------------------------------------------------------------------------------------------
%------------------------------------------------------------------------------------------------------------------
\subsection{Temperature dependence of the breakdown voltage}
\begin{figure}
\centering
\resizebox{\columnwidth}{!}{\includegraphics{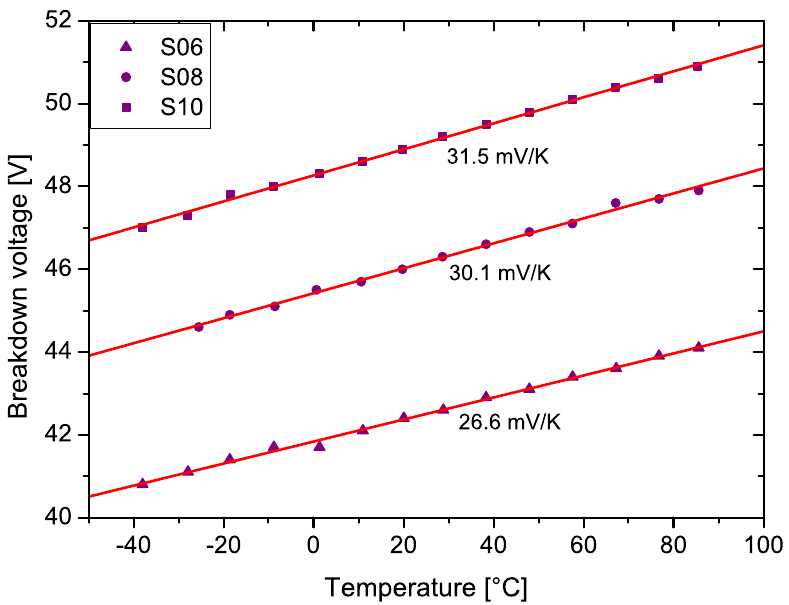}}
\caption{Breakdown voltage vs. temperature. The error bars are within the symbols.}
\label{fig:VBD_vs_Temp}       
\end{figure}

Especially in automotive applications, the systems must operate in a wide temperature range. Independent of whether the photosensor has a temperature stabilization/compensation or not, a low temperature coefficient of the breakdown voltage ($V_{bd}$) is beneficial. For our TAPD, we measure a linear decrease of $V_{bd}$ with temperature (see Fig.~\ref{fig:VBD_vs_Temp}).
The slopes increase with increasing pillar diameter from $\SI{26}{mV/^\circ C}$ for the S06 structure to $\SI{31.5}{mV/^\circ C}$ for the S10 structure. These values are comparable with respect to $\SI{22}{mV/^\circ C}$ for planar KETEK SiPMs \cite{KETEK2019}, despite the clearly larger depletion width and the increased breakdown voltage.
The breakdown voltage was determined from the inverse logarithmic derivative (ILD) of the reverse current-voltage-characteristic with low-level light illumination, as described in \cite{Klanner2017}.
The simulated breakdown voltages (see Tab.~\ref{tab:structures}) overestimate the experimental values at room temperature. The discrepancy decreases from \SI{5}{V} for the S10, to \SI{4.7}{V} for the S08 and to \SI{1}{V} for the S06 structure.
We attribute this observation to the fact that the deviation of the processed structures from a perfect sphere increases with increasing tip size.
For the S10 structure, the shape of the tip is closer to the one of an ellipse than a sphere. Here, the breakdown voltage is defined by the point with the lowest curvature.

\newpage
\section{Conclusion}
\label{sec_Conclusion}
In this work, we presented a new Silicon Photomultiplier, the Tip Avalanche Photodiode. The TAPD concept is based on quasi-spherical electrodes (tips) which are placed inside an epitaxial layer.
We gave a theoretical overview of the physical models and presented the metrological characterization of the existing prototypes.
Our simulations are in a good agreement with the obtained experimental data.
With the TAPD we achieved a record photon detection efficiency over a wide spectral range from \SI{400}{\nano m} to \SI{905}{\nano m} with a small micro-cell pitch of 15 µm. In combination with a high dynamic range, a fast recovery time (\SI{4}{\nano s}), and a low temperature coefficient of the breakdown voltage (\SI{26}{mV/K}), this SiPM is a promising detector for a large variety of applications.

\bibliographystyle{IEEEtran}
\bibliography{bibfile}

\begin{thebibliography}{10}
\providecommand{\url}[1]{#1}
\csname url@samestyle\endcsname
\providecommand{\newblock}{\relax}
\providecommand{\bibinfo}[2]{#2}
\providecommand{\BIBentrySTDinterwordspacing}{\spaceskip=0pt\relax}
\providecommand{\BIBentryALTinterwordstretchfactor}{4}
\providecommand{\BIBentryALTinterwordspacing}{\spaceskip=\fontdimen2\font plus
\BIBentryALTinterwordstretchfactor\fontdimen3\font minus
  \fontdimen4\font\relax}
\providecommand{\BIBforeignlanguage}[2]{{%
\expandafter\ifx\csname l@#1\endcsname\relax
\typeout{** WARNING: IEEEtran.bst: No hyphenation pattern has been}%
\typeout{** loaded for the language `#1'. Using the pattern for}%
\typeout{** the default language instead.}%
\else
\language=\csname l@#1\endcsname
\fi
#2}}
\providecommand{\BIBdecl}{\relax}
\BIBdecl

\bibitem{A}
\BIBentryALTinterwordspacing
{Z. Sadygov }, ``"avalanche detector", ru patent no. 2102820 (1998).''
  [Online]. Available:
  \url{https://www1.fips.ru/fips_servl/fips_servlet?DB=RUPAT\&DocNumber=2102820}
\BIBentrySTDinterwordspacing

\bibitem{B}
\BIBentryALTinterwordspacing
{V. Golovin, M. Tarasov, G. Bondarenko }, ``"avalanche photodetector", ru
  patent no. 2142175 (1998).'' [Online]. Available:
  \url{https://www1.fips.ru/fips_servl/fips_servlet?DB=RUPAT\&DocNumber=2142175}
\BIBentrySTDinterwordspacing

\bibitem{1}
P.~Buzhan \emph{et~al.}, ``An advanced study of silicon photomultiplier,''
  \emph{ICFA Instrum. Bull.}, vol.~23, pp. 28--41, 2001.

\bibitem{2}
K.~{Sato}, K.~{Yamamoto}, K.~{Yamamura}, S.~{Kamakura}, and S.~{Ohsuka},
  ``Application oriented development of multi-pixel photon counter (mppc),'' in
  \emph{IEEE Nuclear Science Symposuim Medical Imaging Conference}, 2010, pp.
  243--245.

\bibitem{C}
A.~G. {Stewart}, V.~{Saveliev}, S.~J. {Bellis}, D.~J. {Herbert}, P.~J.
  {Hughes}, and J.~C. {Jackson}, ``Performance of 1-${\hbox {mm}}^{2}$ silicon
  photomultiplier,'' \emph{IEEE Journal of Quantum Electronics}, vol.~44,
  no.~2, pp. 157--164, 2008.

\bibitem{3}
M.~{Mazzillo}, G.~{Condorelli}, D.~{Sanfilippo}, G.~{Valvo}, B.~{Carbone},
  G.~{Fallica}, S.~{Billotta}, M.~{Belluso}, G.~{Bonanno}, L.~{Cosentino},
  A.~{Pappalardo}, and P.~{Finocchiaro}, ``Silicon photomultiplier technology
  at stmicroelectronics,'' \emph{IEEE Transactions on Nuclear Science},
  vol.~56, no.~4, pp. 2434--2442, 2009.

\bibitem{4}
\BIBentryALTinterwordspacing
C.~Piemonte, ``A new silicon photomultiplier structure for blue light
  detection,'' \emph{Nuclear Instruments and Methods in Physics Research
  Section A: Accelerators, Spectrometers, Detectors and Associated Equipment},
  vol. 568, no.~1, pp. 224 -- 232, 2006, new Developments in Radiation
  Detectors. [Online]. Available:
  \url{http://www.sciencedirect.com/science/article/pii/S016890020601271X}
\BIBentrySTDinterwordspacing

\bibitem{5}
\BIBentryALTinterwordspacing
P.~Bérard, M.~Couture, P.~Deschamps, F.~Laforce, and H.~Dautet,
  ``Characterization study of a new uv–sipm with low dark count rate,''
  \emph{Nuclear Instruments and Methods in Physics Research Section A:
  Accelerators, Spectrometers, Detectors and Associated Equipment}, vol. 695,
  pp. 35 -- 39, 2012, new Developments in Photodetection NDIP11. [Online].
  Available:
  \url{http://www.sciencedirect.com/science/article/pii/S0168900211020821}
\BIBentrySTDinterwordspacing

\bibitem{6}
\BIBentryALTinterwordspacing
C.~Dietzinger, T.~Ganka, W.~Gebauer, N.~Miyakawa, P.~Iskra, and F.~Wiest,
  ``Silicon photomultipliers with enhanced blue-light sensitivity,'' in
  \emph{Talk given at PhotoDet, Seoul, Korea (South)}, 2012. [Online].
  Available:
  \url{https://indico.cern.ch/event/164917/session/0/contribution/18/material/slides/0.pdf}
\BIBentrySTDinterwordspacing

\bibitem{7}
R.~Mirzoyan, P.~Buzhan, B.~Dolgoshein, V.~Kaplin, E.~Popova, and M.~Teshima,
  ``{ SiPM : on the Way at Becoming an Ideal Low Light Level Sensor},'' in
  \emph{Talk given at IEEE Nucl. Sci. Symp. Med. Imaging Conf., Knoxville,
  USA}, 2010.

\bibitem{Vinogradov2011}
S.~Vinogradov, T.~Vinogradova, V.~Shubin, D.~Shushakov, and C.~Sitarsky,
  ``{Efficiency of solid state photomultipliers in photon number resolution},''
  \emph{IEEE Transactions on Nuclear Science}, vol.~58, no. 1 PART 1, pp.
  9--16, 2011.

\bibitem{9}
\BIBentryALTinterwordspacing
F.~Acerbi, G.~Paternoster, A.~Gola, N.~Zorzi, and C.~Piemonte, ``Silicon
  photomultipliers and single-photon avalanche diodes with enhanced nir
  detection efficiency at fbk,'' \emph{Nuclear Instruments and Methods in
  Physics Research Section A: Accelerators, Spectrometers, Detectors and
  Associated Equipment}, vol. 912, pp. 309 -- 314, 2018, new Developments In
  Photodetection 2017. [Online]. Available:
  \url{http://www.sciencedirect.com/science/article/pii/S0168900217313542}
\BIBentrySTDinterwordspacing

\bibitem{Broadcom2020}
\BIBentryALTinterwordspacing
{Product announcement NIR-SiPM, Broadcom}, ``Broadcom’s nir sipm technology
  sets new performance standards for lidar.'' [Online]. Available:
  \url{https://www.broadcom.com/products/optical-sensors/silicon-photomultiplier-sipm}
\BIBentrySTDinterwordspacing

\bibitem{12}
\BIBentryALTinterwordspacing
H.~Sigmund, ``Photoelectrical properties of spherical avalanche diodes in
  silicon,'' \emph{Infrared Physics}, vol.~8, no.~4, pp. 259 -- 264, 1968.
  [Online]. Available:
  \url{http://www.sciencedirect.com/science/article/pii/0020089168900341}
\BIBentrySTDinterwordspacing

\bibitem{G}
\BIBentryALTinterwordspacing
{A. Gasanov, V. Golovin, Z. Sadygov, N. Yusipov }, ``"avalanche semiconductor
  radiation detector", ru patent no. 1702831 (1989).'' [Online]. Available:
  \url{https://www1.fips.ru/fips_servl/fips_servlet?DB=RUPAT\&DocNumber=1702831}
\BIBentrySTDinterwordspacing

\bibitem{13}
Z.~Y. {Sadygov}, I.~M. {Zheleznykh}, N.~A. {Malakhov}, V.~N. {Jejer}, and T.~A.
  {Kirillova}, ``Avalanche semiconductor radiation detectors,'' \emph{IEEE
  Transactions on Nuclear Science}, vol.~43, no.~3, pp. 1009--1013, 1996.

\bibitem{14}
\BIBentryALTinterwordspacing
K.~Linga, Y.~Yevtukhov, and B.~Liang, ``{Near infrared single photon avalanche
  detector with negative feedback and self quenching},'' in \emph{Infrared
  Systems and Photoelectronic Technology IV}, E.~L. Dereniak, J.~P. Hartke,
  P.~D. LeVan, R.~E. Longshore, and A.~K. Sood, Eds., vol. 7419, International
  Society for Optics and Photonics.\hskip 1em plus 0.5em minus 0.4em\relax
  SPIE, 2009, pp. 165 -- 172. [Online]. Available:
  \url{https://doi.org/10.1117/12.826908}
\BIBentrySTDinterwordspacing

\bibitem{H}
\BIBentryALTinterwordspacing
{Z. Ya. Sadygov }, ``“microchannel avalanche photodiode,” ru patent no.
  2316848 (2008).'' [Online]. Available:
  \url{http://www1.fips.ru/fips_servl/fips_servlet?DB=RUPAT\&DocNumber=2316848}
\BIBentrySTDinterwordspacing

\bibitem{15}
\BIBentryALTinterwordspacing
N.~Anfimov, I.~Chirikov-Zorin, A.~Dovlatov, O.~Gavrishchuk, A.~Guskov,
  N.~Khovanskiy, Z.~Krumshtein, R.~Leitner, G.~Meshcheryakov, A.~Nagaytsev,
  A.~Olchevski, T.~Rezinko, A.~Sadovskiy, Z.~Sadygov, I.~Savin, V.~Tchalyshev,
  I.~Tyapkin, G.~Yarygin, and F.~Zerrouk, ``Novel micropixel avalanche
  photodiodes (mapd) with super high pixel density,'' \emph{Nuclear Instruments
  and Methods in Physics Research Section A: Accelerators, Spectrometers,
  Detectors and Associated Equipment}, vol. 628, no.~1, pp. 369 -- 371, 2011,
  vCI 2010. [Online]. Available:
  \url{http://www.sciencedirect.com/science/article/pii/S0168900210015457}
\BIBentrySTDinterwordspacing

\bibitem{Patent_TAPD}
\BIBentryALTinterwordspacing
{P. Iskra, S. Vinogradov }, ``Radiaton detector, method for producing a
  radiation detector and method for operating a radiation detector,'' european
  Patent Application EP3640682A1. [Online]. Available:
  \url{https://patents.google.com/patent/EP3640682A1}
\BIBentrySTDinterwordspacing

\bibitem{JayantBaliga1976}
B.~{Jayant Baliga} and S.~K. Ghandhi, ``{Analytical solutions for the breakdown
  voltage of abrupt cylindrical and spherical junctions},'' \emph{Solid-State
  Electronics}, vol.~19, pp. 739--744, 1976.

\bibitem{Sze2007}
S.~M. Sze and K.~K. NG, \emph{{Physics of Semiconductor Devices}},
  3rd~ed.\hskip 1em plus 0.5em minus 0.4em\relax John Wiley {\&} Sons, Inc,
  2007.

\bibitem{crank_nicolson_1947}
J.~Crank and P.~Nicolson, ``A practical method for numerical evaluation of
  solutions of partial differential equations of the heat-conduction type,''
  \emph{Mathematical Proceedings of the Cambridge Philosophical Society},
  vol.~43, no.~1, p. 50–67, 1947.

\bibitem{VanOverstraeten1970}
R.~{Van Overstraeten} and H.~{De Man}, ``{Measurement of the ionization rates
  in diffused silicon p-n junctions},'' \emph{Solid-State Electronics},
  vol.~13, no.~5, pp. 583--608, 1970.

\bibitem{McINTYRE1966}
R.~J. McIntyre, ``{Multiplication Noise in Uniform Avalanche Diodes},''
  \emph{IEEE Transactions on Electron Devices}, vol. ED-13, no.~1, pp.
  164--168, 1966.

\bibitem{Oide2007}
H.~Oide, H.~Otono, S.~Yamashita, T.~Yoshioka, H.~Hano, and T.~Suehiro,
  \emph{\BIBforeignlanguage{English}{Proceedings of Science}}, 2007,
  international Workshop on New Photon-Detectors, PD 2007 ; Conference date:
  27-06-2007 Through 29-06-2007.

\bibitem{Renker2009}
D.~Renker and E.~Lorenz, ``{"Advances in solid state photon detectors"},''
  \emph{Journal of Instrumentation}, vol.~4, no.~04, pp. P04\,004--P04\,004,
  apr 2009.

\bibitem{Popova2012}
E.~Popova, ``{Evaluation of high UV sensitive SiPMs from MEPhI/MPI for use in
  liquid argon },'' \emph{PoS}, vol. PhotoDet 2012, p. 034, 2013.

\bibitem{RB_Series2020}
{Product Data Sheet RB-Series SiPM, On Semiconductor}, online available at:
  \url { https://www.onsemi.com/pub/Collateral/MICRORB-SERIES-D.PDF}.

\bibitem{Piemonte2012}
C.~{Piemonte}, A.~{Ferri}, A.~{Gola}, A.~{Picciotto}, T.~{Pro}, N.~{Serra},
  A.~{Tarolli}, and N.~{Zorzi}, ``Development of an automatic procedure for the
  characterization of silicon photomultipliers,'' in \emph{2012 IEEE Nuclear
  Science Symposium and Medical Imaging Conference Record (NSS/MIC)}, 2012, pp.
  428--432.

\bibitem{Vachon2018}
V.~Chaumat, ``Sipm pde measurement with continuous and pulsed light,''
  \emph{PhotoDet 2012}, p. 058, 2013.

\bibitem{Vinogradov2016}
S.~{Vinogradov}, ``Precise metrology of sipm: Measurement and reconstruction of
  time distributions of single photon detections and correlated events,'' in
  \emph{2016 IEEE Nuclear Science Symposium, Medical Imaging Conference and
  Room-Temperature Semiconductor Detector Workshop (NSS/MIC/RTSD)}, 2016, pp.
  1--4.

\bibitem{Engelmann2018}
{E. Engelmann}, ``Sipm noise measurement with waveform analysis,'' in
  \emph{Talk given at International Conference on the Advancement of Silicon
  Photomultipliers (ICASiPM), Schwetzingen, Germany}, 2018.

\bibitem{Thorlabs2019}
{Calibrated Si Photodiode - FDS100-CAL from Thorlabs}, online available at:
  \url {https://www.thorlabs.com/thorproduct.cfm?partnumber=FDS100-CAL}.

\bibitem{Otte2017}
\BIBentryALTinterwordspacing
A.~N. Otte, D.~Garcia, T.~Nguyen, and D.~Purushotham, ``Characterization of
  three high efficiency and blue sensitive silicon photomultipliers,''
  \emph{Nuclear Instruments and Methods in Physics Research Section A:
  Accelerators, Spectrometers, Detectors and Associated Equipment}, vol. 846,
  pp. 106 -- 125, 2017. [Online]. Available:
  \url{http://www.sciencedirect.com/science/article/pii/S0168900216309901}
\BIBentrySTDinterwordspacing

\bibitem{Gola2019}
A.~Gola, F.~Acerbi, M.~Capasso, M.~Marcante, A.~Mazzi, G.~Paternoster,
  C.~Piemonte, V.~Regazzoni, and N.~Zorzi, ``Nuv-sensitive silicon
  photomultiplier technologies developed at fondazione bruno kessler,''
  \emph{Sensors}, vol.~19, p. 308, 01 2019.

\bibitem{Yamamoto2018}
{K. Yamamoto}, ``Recent development of mppc at hamamatsu for photon counting
  applications,'' in \emph{Talk given at5th International Workshop on New
  Photon Detectors (PD18), Tokyo, Japan}, 2018.

\bibitem{Eckert2010}
P.~Eckert, H.-C. Schultz-Coulon, W.~Shen, R.~Stamen, and A.~Tadday,
  ``{"Characterisation studies of silicon photomultipliers"},'' \emph{Nuclear
  Instruments and Methods in Physics Research Section A: Accelerators,
  Spectrometers, Detectors and Associated Equipment}, vol. 620, no.~2, pp. 217
  -- 226, 2010.

\bibitem{Engelmann_PhD_2018}
E.~Engelmann, ``Dark count rate of silicon photomultipliers - metrological
  characterization and suppression,'' {Cuvillier, Göttingen, 2018}.

\bibitem{KETEK2019}
{Product Data Sheet PM1125-WB-B0}, online available at: \url {
  https://www.ketek.net/wp-content/uploads/KETEK-PM1125-WB-B0-Datasheet.pdf}.

\bibitem{ONeill2015}
K.~O'Neill and C.~Jackson, ``{"SensL B-Series and C-Series silicon
  photomultipliers for time-of-flight positron emission tomography"},''
  \emph{Nuclear Instruments and Methods in Physics Research, Section A:
  Accelerators, Spectrometers, Detectors and Associated Equipment}, vol. 787,
  pp. 169--172, 2015.

\bibitem{Pagano2012}
R.~Pagano, D.~Corso, S.~Lombardo, G.~Valvo, D.~N. Sanfilippo, G.~Fallica, and
  S.~Libertino, ``{"Dark Current in Silicon Photomultiplier Pixels: Data and
  Model"},'' \emph{IEEE Transactions on Electron Devices}, vol.~59, no.~9, pp.
  2410--2416, 2012.

\bibitem{Klanner2019}
\BIBentryALTinterwordspacing
R.~Klanner, ``Characterisation of sipms,'' \emph{Nuclear Instruments and
  Methods in Physics Research Section A: Accelerators, Spectrometers, Detectors
  and Associated Equipment}, vol. 926, pp. 36 -- 56, 2019, silicon
  Photomultipliers: Technology, Characterisation and Applications. [Online].
  Available:
  \url{http://www.sciencedirect.com/science/article/pii/S0168900218317091}
\BIBentrySTDinterwordspacing

\bibitem{Klanner2017}
\BIBentryALTinterwordspacing
V.~Chmill, E.~Garutti, R.~Klanner, M.~Nitschke, and J.~Schwandt, ``Study of the
  breakdown voltage of sipms,'' \emph{Nuclear Instruments and Methods in
  Physics Research Section A: Accelerators, Spectrometers, Detectors and
  Associated Equipment}, vol. 845, pp. 56 -- 59, 2017, proceedings of the
  Vienna Conference on Instrumentation 2016. [Online]. Available:
  \url{http://www.sciencedirect.com/science/article/pii/S016890021630256X}
\BIBentrySTDinterwordspacing

\end{thebibliography}

\end{document}